\documentclass[cits]{PoS}
\usepackage[numbers,square]{natbib}
\usepackage[T1]{fontenc}
\usepackage[varg]{txfonts}

\def\aj{AJ}%
\def\araa{ARA\&A}%
\def\apj{ApJ}%
\def\apjl{ApJ}%
\def\apjs{ApJS}%
\def\apss{Ap\&SS}%
\def\aap{A\&A}%
\def\mnras{MNRAS}%
\def\pasp{PASP}%
\def\nat{Nature}%
\title{Thermonuclear Supernovae}

\ShortTitle{Thermonuclear Supernovae}

\author{\speaker{F.~K.~R{\"o}pke}\\
        Max-Planck-Institut f{\"u}r Astrophysik,
        Karl-Schwarzschild-Str.~1, D-85741 Garching, Germany\\
        E-mail: \email{fritz@mpa-garching.mpg.de}}

\abstract{The application of Type Ia supernovae (SNe~Ia) as distance
          indicators in cosmology calls for a sound understanding of
          these objects. Recent years have seen a brisk development of
          astrophysical models which explain SNe~Ia as
          thermonuclear explosions of white dwarf stars. While the
          evolution of the progenitor is still uncertain, the
          explosion mechanism certainly involves the propagation of a
          thermonuclear flame through the white dwarf
          star. Three-dimensional hydrodynamical simulations allowed
          to study a wide variety of possibilities involving subsonic
          flame propagation (deflagrations), flames accelerated by
          turbulence, and supersonic detonations. These possibilities
          lead to a variety of scenarios. I review the currently
          discussed  approaches and present some recent results from
          simulations of the turbulent deflagration model and the
          delayed detonation model.}

\FullConference{Supernovae: lights in the darkness (XXIII Trobades Cient\'ifiques de la Mediterr\`ania)\\
                 October 3-5 2007\\
                 Mao, Menorca, Spain}

\begin{document}

\section{Introduction}

Thermonuclear supernova explosions are an astrophysical model for the
astronomical class of Type Ia supernovae (SNe~Ia henceforth). These objects
are of interest in many fields of astrophysics and astronomy. Being
one of the main sources
of iron group elements, SNe~Ia contribute to the chemical
evolution of galaxies (e.g., \citep{francois2004a}). They affect star
formation and drive shock waves in the interstellar and intergalactic
media.

The main driver of SN~Ia research over the past years, however, has
been their application in observational
cosmology. Here, SNe~Ia were employed as distance indicators
(as put forward by \citep{branch1992a}). At redshifts above 0.5, a
significant deviation from 
the linear Hubble law was noticed which led to the spectacular 
interpretation of the Universe currently undergoing
an accelerated expansion
\citep{riess1998a,perlmutter1999a}. The determination of the force
driving this acceleration 
is perhaps one of the greatest challenges in contemporary physics. For
the time being,
it is parametrized as ``dark energy'' (e.g., \citep{leibundgut2001a}).
The simplest form of dark energy is a
cosmological constant, but more complicated contributions to the
energy-momentum tensor in the Einstein equations are also
conceivable. A first step to determine the nature of dark energy
would be to constrain its equation of state. SNe~Ia seem to be
a suitable tool for this task and currently two major campaigns
\citep{astier2005a, wood-vasey2007a} apply them in distance
determinations of hundreds of supernovae out to redshifts of $z \sim 1$.  
The large number of observations is necessary to reduce the
statistical errors because putting tight constraints on dark energy
equation of state requires a high accuracy of the distance
determinations. 

From a theoretical point of view, however, the applicability of SNe~Ia as distance indicators
is still not satisfactorily answered. SNe~Ia are remarkably uniform
events by astrophysical standards, but evidently no
standard candles. Only a calibration of the distance measurements
according to empirical correlations between observables provides the
necessary accuracy for the determination of cosmological parameters. 
Such calibrations may be afflicted with systematic errors. Being
derived from a set of nearby well-observed SNe~Ia, there is no
guarantee that they perform well for supernovae at high
redshifts, too. 
Getting a handle on these uncertainties is one of the goals of modeling
SNe~Ia.

\section{Astrophysical modeling}

The cornerstones of the astrophysical model of SNe~Ia derive from the
fundamental characteristics of these events:
\begin{itemize}
\item Evidently, SNe
Ia belong to the most energetic cosmic explosions, releasing about
$10^{51} \, \mathrm{erg}$ of energy. For a short period of time
they can outshine an entire galaxy consisting of tens of billions of
stars.
\item
SNe~Ia spectra are characterized by lacking
indications of hydrogen and helium which together with a pronounced P
Cygni silicon line at maximum light classifies these objects
\citep{wheeler1990a}. Lines
of intermediate-mass elements (IME, such as Si, Ca, Mg and S) and
oxygen are observed in near-maximum light spectra
(e.g., \citep{filippenko1997a}).
\item SNe~Ia form a class of remarkable
homogeneity with respect to observed lightcurves and spectra
(e.g., \citep{branch1992a}). 
\end{itemize}

Assuming supernovae to generally originate from single stellar objects, only
their gravitational binding energy, released in a collapse towards a
compact object \citep{zwicky1938a}, or its nuclear energy, released in
explosive 
reactions \citep{hoyle1960a}, come into consideration as possible
energy sources. In the 
particular case of SNe~Ia, no compact object is found in the remnants
excluding the first possibility. The homogeneity of the class of SNe~Ia
and the fact that no hydrogen is found in
their spectra provides a strong hint that the object
undergoing the nuclear explosion may be a white dwarf (WD) star consisting
of carbon and oxygen (C+O).

Lightcurves of SNe~Ia rise over a time scale of several days and
decline over months. It is therefore clear that they cannot be powered
directly by the explosion since the temperatures fall off much too rapidly in the
expansion. This problem was solved by noting that the $^{56}$Ni produced in large
amounts in the explosive thermonuclear burning provides the energy
source for the optical event by radioactive decay to $^{56}$Co and
$^{56}$Fe \citep{truran1967a, colgate1969a}.

\subsection{Progenitor evolution and ignition}
\label{sect:progenitor_ignition}

A single WD is an inert object. How can it reach an explosive state?
The only way to introduce the necessary dynamics into the system is to
assume it to be part of a binary system and to gain matter from the
companion. Several models have been proposed for this progenitor
evolution. Here, we will discuss only models resulting
from the so-called \emph{single-degenerate Chandrasekhar-mass
  scenario}, which has received by far most attention recently. For
a summary of alternative models, we refer to \citep{hillebrandt2000a}.

In the \emph{single degenerate} scenario
\citep{whelan1973a,nomoto1982a,iben1984a}, the WD accretes matter from
a non-degenerate companion (either a main sequence star, or an AGB
star). 
This idea was recently supported by the detection of the potential
companion of Tycho Brahe's supernova of 1572, which is a solar-type star \citep{ruiz-lapuente2004a}.
By accreting material from this companion and burning it
hydrostatically at the surface, the WD may reach the
Chandrasekhar mass $M_\mathrm{Ch}$.
Limiting the fuel available in the explosion to
$M_\mathrm{Ch} \sim 1.4\, \mathrm{M}_{\odot}$, this \emph{Chandrasekhar-mass
  model} appears particularly 
favorable since it provides a natural explanation for the striking
uniformity of SNe~Ia in the gross observational features. At the same time,
it is afflicted with great uncertainties. Achieving a stable
mass transfer in the progenitor binary system to build up a
Chandrasekhar mass WD is highly non-trivial
(e.g., \citep{nomoto1985a}) and the observational
evidence for such systems is sparse.

When the WD  approaches
the Chandrasekhar limit,
the density at the center of the WD increases rapidly so
that fusion of carbon ignites. Contrary to the situation in main
sequence stars, the degenerate material of the WD does not allow for
moderation of the burning by expansion. Heat transport is achieved
here by convection giving rise to a stage of
convective carbon burning that lasts for several hundred years. This
phase is terminated by one or more small spatial regions undergoing a 
thermonuclear runaway, marking the birth of a thermonuclear flame
and the onset of the explosion. 
The convective burning stage and
the conditions at flame ignition are extremely hard to model both
analytically and numerically. Therefore the exact shape and
location of the first flame spark(s) is not yet well
constrained \citep{garcia1995a,woosley2004a, iapichino2006a,
  kuhlen2006a, hoeflich2002a}. 
Does ignition occur in a single spot or in multiple sparks with a stochastic
distribution? Does it take place at the center of the WD or do
pre-ignition convective motions lead to large asymmetries?
Evidently, the ignition structure is a  crucial initial
parameter in multi-dimensional explosion models \citep{roepke2006a,
  schmidt2006a,plewa2004a,roepke2007a, sim2007a, hillebrandt2007a}.

\subsection{Flame propagation and explosion}
\label{sect:flame_explosion}

The goal of SN~Ia explosion models is to follow the propagation of the
thermonuclear flame from its ignition near the center of the WD
outwards and to determine the nuclear energy release and the
structure of the ejected material.

Hydrodynamics allows for two distinct modes of flame propagation. 
One is the subsonic deflagration in
which the flame is mediated by the thermal conduction of the
degenerate electron gas and the other is a supersonic detonation in
which the burning front is driven by shock waves.
Either one of these modes or a combination of both have been suggested
in different explosion models:

\begin{itemize}
\item \emph{The prompt detonation model} \citep{arnett1969a} attempts
  to explain SNe~Ia by a detonation ignited at the center of the WD
  and propagating 
  outward. This
  produces enough energy for a SN~Ia event. However, ahead of
  a supersonic detonation wave, the fuel cannot expand and is therefore
  incinerated at the high densities of an equilibrium white
  dwarf. This results in the almost complete conversion of the
  material to nickel-peaked nuclear statistical equilibrium
  \citep{arnett1969a}, which is in 
  conflict with the intermediate mass elements observed in SN~Ia
  spectra. Such nucleosynthetic problems rule out pure
  detonations as a standard model for SN~Ia explosions.

\item \emph{The deflagration model}
  \citep{nomoto1976a} assumes the flame to
  propagate in the subsonic deflagration mode. The laminar burning speed of the
  deflagration flame is determined by microphysical transport
  processes. For conditions of carbon burning in C+O WDs it is highly
  subsonic \citep{timmes1992a} and therefore the flame propagates
  far too slowly to explain SN~Ia explosions. The expansion of the star
  quenches burning before the WD gets unbound. On the other hand, this
  model can cure the problem of nucleosynthesis, since rarefaction
  waves travel ahead of the flame with sound speed and lower the
  fuel density prior to burning. Thus, the material can partly be processed into
  intermediate mass elements.
  The deflagration model undergoes a significant improvement when multidimensional
  effects are taken into account as will be discussed in detail in
  Sect.~\ref{sect:defl}.

\item \emph{The delayed detonation model} \citep{ivanova1974a,
  khokhlov1991a, woosley1994a} unites the
  advantages of the deflagration and the detonation models.
  Burning
  starts out in the slow deflagration mode
  pre-expanding the star. At some point, a transition from the initial
  subsonic deflagration to a supersonic detonation takes place.
  This detonation is an easy way to explain the energy release
  necessary for a SN~Ia explosion. The
  important notion in this model is that a detonation in low density
  fuel (pre-expanded in the deflagration stage)
  can lead to only partial burning and is therefore capable
  of generating intermediate mass elements. A detailed account of
  this model will be given in Sect.~\ref{sect:dd}.

\item \emph{The pulsational delayed detonation model}
  \citep{arnett1994b} is
  similar to the delayed detonation model in the sense that it combines an initial
  deflagration with a later detonation. The flame is assumed to
  propagate in the initial deflagration phase with its laminar burning
  speed and pre-expands the star. Due to the slow flame velocity, the
  burning front stalls and fails to unbind the star. The WD then
  re-contracts giving the interface between burnt and unburnt material
  enough time to mix and to become nearly isothermal. Compressional
  heating finally triggers a detonation at densities that are lower
  than that prior to the first expansion phase. 
  The assumption
  of the flame propagating with the pure laminar burning velocity in
  the deflagration phase originally made in this model seems
  unrealistic because of the flame 
  instabilities and the resulting turbulent flame acceleration. Recent
  multidimensional deflagration models
  \citep{reinecke2002d,gamezo2003a} demonstrated that taking  these
  effects into
  account, the star is likely to get unbound instead of
  recontracting. 
\end{itemize}

\section{Numerical modeling}

Numerical models of SN~Ia explosions have to face three major
challenges. The vast range of relevant length scales cannot be
resolved in computational models in the foreseeable future.
It extends from the radius of the WD star
($\sim$$2000\,\mathrm{km}$ at the onset of the explosion and expanding
in the process) down to the flame width which is well below one
centimeter and, including turbulence effects, to the Kolmogorov scale
of less than a millimeter.
Apart from this problem, numerical models need
to take into account inherently three-dimensional physical phenomena
and to solve the equations of nuclear burning. To meet all these
requirements in a 
single simulation will be impossible in the foreseeable
future. 
Therefore the problem has been tackled in different
approaches. 

The first path towards SN~Ia explosion modeling is to
restrict the simulations to only one spatial dimension. Here, in
principle a resolution of the relevant scales is achievable and a
detailed description of the nuclear reactions is feasible. 
However, this approach fails to consistently incorporate crucial
three-dimensional physical mechanisms.

In multi-dimensional simulations (see \citep{roepke2006c,roepke2006d}
for recent reviews), the computational
costs of modeling the explosion hydrodynamics is prohibitive to
directly resolve all relevant scales as well as details of the nuclear
processes. While the latter may be improved in the forthcoming years,
and is meanwhile separated from the actual explosion simulations
still maintaining a reasonable accuracy (for an approach based on
reconstructing the nucleosynthesis from tracer particles, see
\citep{travaglio2004a,roepke2006b}), even a drastic
increase in computational capabilities will not allow for a resolution
of all relevant scales in multi-dimensional simulations. Consequently,
additional modeling effort is required in order to implement a
consistent description of flame propagation in these simulations (see
Sect.~\ref{sect:defl}). 

A third approach is to study specific effects on a limited range of
spatial scales, in order to validate assumptions and improve modeling
techniques of the large-scale SN~Ia simulations.

\subsection{Objectives of modeling Type Ia supernovae}

Three-dimensional numerical models of Type Ia supernovae strive for a
self-consistent astrophysical description of these astronomical
events. The goal is to achieve a viable model by starting out from
fundamental physical laws (``first principles'') and to formulate the
model without introducing free parameters. Avoiding tunable parameters
allows the simulations to gain a high level of predictive power. Only
this way, a direct comparison of synthetic observables derived from
the simulations with actual SN~Ia observations facilitates a thorough
validation of the modeling approach.

This aim of multidimensional models is significantly different from
earlier one-dimensional modeling approaches. Here, the flame
propagation velocity, which is determined by inherently
multidimensional effects such as instabilities and turbulences, cannot
be consistently determined. It thus introduces a free parameter which
has significant impact on the result of the simulated explosion. With
such a powerful tunable parameter an impressive agreement between
one-dimensional simulations and observations could be achieved. This
in turn tells us a great deal about the global properties of the
explosion mechanism. Successful one-dimensional models (e.g.\ W7
\citep{nomoto1984a}) bring forward the
average flame velocity that needs to be attained in different stages
of the explosion and remain the benchmark of the (spherically
averaged) chemical composition of the explosion ejecta.

Successful multidimensional models, however, would diverge from this
way of tackling SN~Ia modeling as an inverse problem. Modeling the
explosion from first principles, they make predictions of
observables. This paves the way to investigating questions related to
the cosmological use of SNe~Ia: What is the precision of SNe~Ia as
distance indicators? What is the reason for the SN~Ia diversity? How
does the diversity relate to variations in the properties of the
progenitor system? How do these variations translate (via the explosion
process) to observable features? How can one improve the calibration
techniques of SN~Ia cosmology? 
With ongoing observational campaigns, projects currently starting and
future surveys, the number of observed SNe~Ia will increase from
currently a few hundred per year to several thousands. This calls for
a sound theoretical understanding of SNe~Ia as an astrophysical
phenomenon. A rapid development of the field of multidimensional SN~Ia
simulations gives hope that these models will be able to tackle the
questions raised above in the not too far future. 

\subsection{Current status of SN~Ia modeling}

The zoo of variants of the Chandrasekhar-mass
single-degenerate SN~Ia model recently studied in multidimensional
simulations comprizes
\begin{itemize}
\item the \emph{turbulent deflagration model} (see also
  Sect.~\ref{sect:defl}), which is so-far best
  studied in multidimensional simulations \citep{reinecke2002d, reinecke2002b, gamezo2003a, roepke2005a,
    roepke2005b, roepke2006a, schmidt2006a, roepke2007a, roepke2007c}. The advantage of this model
  is that the explosion process can be formulated free of tunable
  parameters. Therefore it is possible to compare the outcome of
  simulations directly to observations in order to assess the validity
  of the model. Systematic tests of the initial parameters of the
  model on the outcome of the exploson process \citep{roepke2004a, roepke2006b} and
  their implication for SN~Ia cosmology \citep{roepke2006e} are possible.
\item the \emph{delayed detonation model} \citep{gamezo2005a,
    golombek2005a, roepke2007b, bravo2007a}, which mainly suffers from
  the uncertain mechanism providing a deflagration-to-detonation
  transition in thermonuclear supernovae. Fixing this uncertain
  parameter to a physically motivated hypothesis, however, the results
  of corresponding numerical simulations look promising (see Sect.~\ref{sect:dd}).
\item the \emph{gravitationally confined detonation model}, which
  arises from asymmetric deflagration flame ignitions. It has been
  claimed that in some cases the deflagration may fail to unbind the
  WD \citep{plewa2004a}. As a
  consequence of the off-center ignition, the ashes of the
  deflagration burning quickly float towards the surface of the WD. If
  still gravitationally bound, they start to sweep around the unburnt
  core of the star and collide on the far side. This collision has
  been claimed to trigger a detonation wave \citep{plewa2004a,
    plewa2007a, townsley2007a} (see,
  however, \citep{roepke2007a,roepke2006f}) propagating inwards and
  burning out the core of the WD. Very energetic and bright events are
  expected from such a model.  
\item the \emph{pulsational reverse detonation model}, which follows the pulsational phase of the
  gravitationally bound WD if no detonation is triggered by the
  collision of ashes as in the gravitationally confined detonation
  scenario. Pulsational contraction has been suggested to trigger a
  detonation reviving the explosion in this case \citep{bravo2006a}. 
\end{itemize}

The goal of a fully self-consistent SN~Ia model that agrees with all
observatonal features is not reached yet. Some of the modeling
approaches, however, seem very promising.
The following part of the paper focuses on a detailed description of
the first two of the above listed models.

\section{The turbulent deflagration model}
\label{sect:defl}

The \emph{turbulent} deflagration model extends the laminar
deflagration model by considering effects of instabilities and
turbulence on the flame propagation. These lead to a significant
acceleration of the burning and to viable explosions of the WD star. 

\subsection{Turbulent deflagratlion in thermonuclear supernovae}

The major effect accelerating the
flame is due to the buoyancy unstable
flame propagation from the center of the star outwards. It leaves
behind light and hot ashes below
the dense fuel -- a density statification inverse to the
gravitational acceleration. In its non-linear stage, the
Rayleigh--Taylor instability leads to the formation of
mushroom-shaped burning bubbles raising into the fuel. The Reynolds
number typical for this situation is as high as $10^{14}$. Clearly,
shear (Kelvin--Helmholtz) instabilities at the interfaces of these
bubbles will generate turbulent eddies which then decay to smaller
scales forming a turbulent energy cascade. The flame will interact
with these eddies down to the Gibson-scale at which the turbulent
velocity fluctuations become comparable to the laminar flame
speed. Below the Gibson scale, the flame burns faster through
turbulent eddies than they can deform it, and the flame propagation
is thus unaffected by turbulence there. This interaction
corrugates the flame increasing its surface, enhancing the net burning
rate, and, consequently, accelerating the effective flame propagation speed.

In the numerical implementation of the deflagration model, the relevance of turbulent
effects amplifies the scale problem since the turbulent
cascade extends to the extremely small Kolmogorov scale ($< 1 \, \mathrm{mm}$) where
the turbulent energy is dissipated into heat. The flame
interaction with the turbulent cascade down to the Gibson scale
must be taken into account.
Current 3D simulations capturing the entire star reach
resolutions 
around one kilometer while the Gibson scale is of the order of
$10^4\,\mathrm{cm}$ at the beginning of the explosion and decreases
steadily.
For large-scale multi-dimensional SN~Ia simulations this has three consequences:
\begin{enumerate}
\item \label{enum2} The internal flame structure cannot be
  resolved. Thus, an effective flame model has to be applied and
  complementary small scale simulations are required. 
\item \label{enum1} It is not possible to fully resolve the interaction of the flame
  with turbulence. Therefore modeling of the effects on unresolved
  scales is necessary.
\item Assumptions about the flame properties on unresolved scales
  \citep{roepke2003a,roepke2004a,roepke2004b,zingale2005a,schmidt2005e} have to be validated in
  separate small-scale simulations. 
\end{enumerate}
Multi-dimensional models are inevitable to consistently determine the
turbulent flame propagation velocity.  Given the wide range of scales
on which the 
flame is affected by turbulence, this is an ambitious project,
additionally challenged by the lack of resolution of the thermonuclear
flame structure. For both problems, different approaches have been
taken, guided by the theory of turbulent
combustion in terrestrial flames (see \citep{peters2000a}).

\subsection{Flame model}

Two major strategies to tackle the problem of the unresolved
internal flame structure can be distinguished. 
On the one hand, a flame capturing
technique \citep{khokhlov1993a} mimics
flame propagation by an artificial 
diffusion mechanism which broadens the internal flame structure to a
certain number
of computational grid cells. It has been applied in SN~Ia
explosion simulations
\citep{gamezo2003a,gamezo2005a, calder2004a,
plewa2004a, townsley2007a}. 

On the other hand, a completely different approach
\citep{reinecke1999a} treats the
flame as a sharp discontinuity 
separating the fuel from the ashes. It is numerically represented
applying the level-set technique \citep{osher1988a}. Here,
the flame front is associated with the zero level set of a scalar
function $G$ representing the distance from the interface. 
A model for flame propagation based on this technique
\citep{smiljanovski1997a} was modified for thermonuclear
flames in SN~Ia explosions \citep{reinecke1999a, hillebrandt2005a}. This
scheme was applied in a number of simulations
\citep{reinecke2002b, reinecke2002d,
  roepke2004c,roepke2005a,roepke2005b, roepke2005c, schmidt2006c,
  schmidt2006a, roepke2006b, roepke2006a, roepke2007a,roepke2007c}, and
the simulations presented below are based on it.

\subsection{Turbulent combustion model}
\label{sect:turb_comb}

The wide range of scales involved in turbulent combustion in
thermonuclear supernovae makes direct simulations virtually
impossible.
Only parts of the flame/turbulence interaction range and of
the resulting flame surface enlargement can be
resolved on the computational grid. This deficit is usually
compensated by attributing an effective 
\emph{turbulent flame speed} $s_\mathrm{t}$ to the unresolved flame
front, which must be determined by theoretical considerations.

One of the cornerstones of the theoretical description of turbulent
combustion is the notion of different regimes of flame/turbulence
interaction \citep{niemeyer1997d}. These regimes are distinguished by
the ability of
turbulent eddies to penetrate the internal flame structure. Since the
Gibson scale is much larger than the flame width for most parts of the
SN~Ia explosion, this will not be the case here and accordingly the
combustion falls into the regime of \emph{wrinkled and corrugated
  flamelets} (e.g., \citep{peters2000a}). In this regime, the full flame structure is corrugated by the
interaction with turbulence and the resulting surface enlargement
accelerates its propagation. The flame propagation
completely decouples from the microphysics of the burning for
sufficiently strong turbulence \citep{damkoehler1940a}. It is
entirely determined by the turbulent velocity fluctuations, that is,
the effective turbulent flame speed $s_\mathrm{t}$ is proportional to the turbulent velocity
fluctuations $v'$. 

One of the challenges of deflagration models of SN~Ia
explosions is thus to determine these velocity fluctuations
correctly. Since the 
resolution in multi-dimensional simulations is insufficient to resolve
the phenomena directly, modeling approaches have to be taken. The
assumption of the flame propagation being entirely driven by buoyancy
instabilities \citep{gamezo2003a,calder2004a,plewa2004a,townsley2007a}
falls short of reproducing the effects of a turbulent cascade which
will develop due to large-scale shear instabilities and dominate the
turbulence properties at small scales. Such effects are taken into
accunt only with appropriate subgrid-scale turbulence models
\citep{niemeyer1995a,schmidt2006c}. Guided by the technique of
\emph{Large Eddy Simulations (LES)},
this model
determines the turbulence energy on unresolved scales based on
conservation laws. It is applied in the simulations discussed below.

In the very late stages of the SN~Ia explosion, the fuel
density drops  due to the
expansion of the WD to values where the flame width becomes broader than the
Gibson length. Then, turbulence penetrates the internal structure of
the flame and it enters the regime of \emph{distributed
  burning} \citep{niemeyer1997d}. Including this burning stage into
thermonuclear supernova simulations \citep{roepke2005a,schmidt2007a}
affects the latest stages of deflagration burning.

\subsection{Example: a highly resolved simulation}

\begin{figure}[t!]
\includegraphics[width = \textwidth]{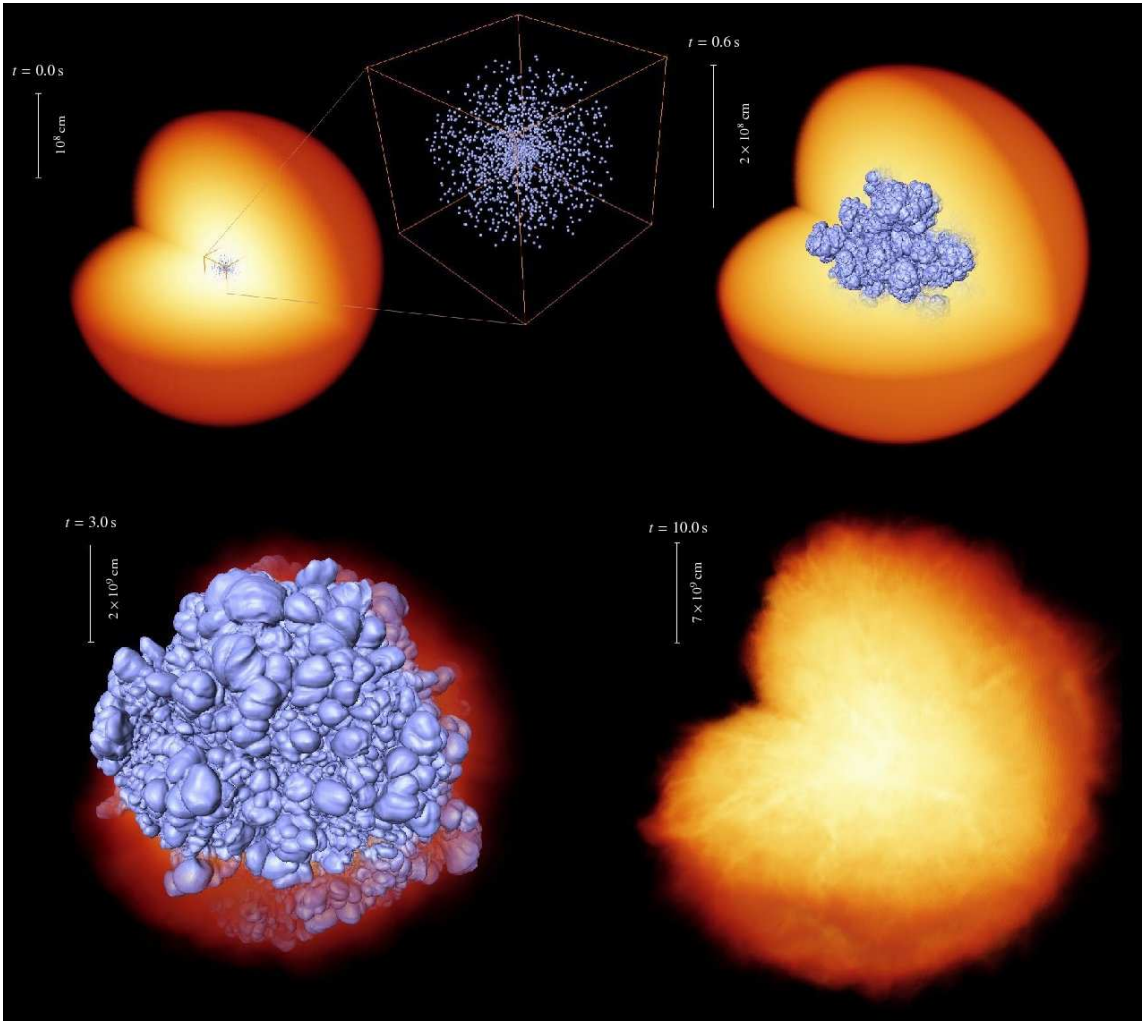}
\caption{Snapshots from a full-star SN~Ia simulation starting from
  a multi-spot ignition scenario. The logarithm of the density is
  volume rendered indicating the extend of the WD star and the
  isosurface corresponds to the thermonuclear flame. The last snapshot
  corresponds to the end of the simulation and is not on scale with
  the earlier snapshots (from \citep{roepke2007c}).\label{fig:evo}}
\end{figure}

Several simulations based on the implementation described above, both
in two and three spatial dimensions, have
been presented \citep{reinecke1999b,reinecke2002b,reinecke2002d}. In the
2D simulations, numerical convergence in the global
quantities was demonstrated. For the implementation on a co-expanding computational
grid, a similar result was found \citep{roepke2005c}. The numerical
convergence naturally arises from
the interplay of the resolved flame front representation with the
turbulent subgrid-scale model. Ideally, a lack of resolution of
large-scale features in the flame front representation should be
compensated by an increased turbulent flame propagation velocity
determined from the subgrid-scale approach. Of course, a certain
threshold of resolution will need to be exceeded in order to reach this regime
in the numerical implementation. A consistent and reliable sub-grid
scale modeling was the goal of a recent highly resolved simulation
\citep{roepke2007c}. This model shall be described here in order to
illustrate the typical flame evolution in the deflagration scenarios
of thermonuclear supernovae.
A full-star \citep{roepke2005b} multi-spot
ignition \citep{roepke2006a} model was set up on a
computational grid comprizing $1024\times 1024$ cells. In combination
with the nested moving grid approach \citep{roepke2006a} this
facilitated an extremely fine-structured initial flame geometry (see
the upper left panel of Figure~\ref{fig:evo}).

Starting from this
initial flame configuration, the
evolution of the flame front in the explosion process is illustrated
by snapshots of the $G=0$ isosurface at $t = 0.6 \, \mathrm{s}$ and $t =
3.0 \, \mathrm{s}$ in
Figure~\ref{fig:evo} (upper right and lower left panels). 
The development of
the flame shape from ignition to $t = 0.6 \, \mathrm{s}$ is
characterized by the formation of the well-known ``mushroom-like'' structures resulting
from buoyancy. 
During the flame evolution, inner structures of smaller
scales catch up with the outer ``mushrooms'' and the initially separated structures
merge forming a connected configuration (see snapshot at $t = 0.6 \,
\mathrm{s}$ of Fig.~\ref{fig:evo}). 
The
continued development of substructure and the merger of features create a
deflagration structure with a complex pattern. Burning and flotation drive
the flame towards the surface of the WD. The fuel density drops as the
flame moves outwards due to radial stratification and the overall
expansion of the WD caused 
by the energy deposit from nuclear burning. Once the fuel density
falls below the threshold for the production of intermediate-mass
elements, nuclear burning ceases. This occurs first
at the leading features of the flame and subsequently in more central
flame regions. Finally, no burning takes place anymore. At this point
the outer ash features have reached the surface layers of the
ejecta (lower left panel of Fig.~\ref{fig:evo}). The following
seconds in the evolution are characterized by hydrodynamical
relaxation towards homologous expansion. This stage is reached to good
approximation at about $10 \, \mathrm{s}$ after ignition \citep{roepke2005c}.

This three-dimensional evolution leads to a remnant of the explosion
with characteristic properties. The density structure has patterns from
unstable and turbulent flame propagation imprinted on it (see the lower
right panel of Fig.~\ref{fig:evo}), and ash regions extend to 
the outermost layers of the expanding cloud of gas.

\subsection{Comparison with observations}
\label{sect:compare}

Due to recent progress in
deriving observables from multi-dimensional deflagration simulations, a direct
comparison with details of observations of nearby
SNe~Ia has come into reach. 
Since the simulations contain no other parameters than the initial
conditions, the question arises of whether they meet observational constraints. Such
constraints result from the global characteristics derived from
observations, observed lightcurves, and spectra taken from nearby SNe
Ia. We will present a comparison of the highly resolved simulation
described above with observational expectations in the following.

The global characteristics derived from SN~Ia observations state that
a valid explosion model should release $\sim$$10^{51}\,\mathrm{erg}$
of energy and produce about $0.4\ldots 0.7\, \mathrm{M}_{\odot}$ of $^{56}$Ni in
the nuclear burning \citep{contardo2000a,stritzinger2006a}. However, there exists a large diversity in the
observations ranging from the class of sub-luminous SNe~Ia (like
SN~1991bg with probably $\sim$$0.1 \, \mathrm{M}_{\odot}$ of $^{56}$Ni) to
super-luminous events (e.g.\ SN~1991T with a $^{56}$Ni mass close to
$1 \, \mathrm{M}_{\odot}$). 
The simulation under consideration here led to an asymptotic kinetic
energy of the ejecta of
$\sim$$8.1 \times 10^{50}\,\mathrm{erg}$ and produced $\sim 0.6 \, \mathrm{M}_{\odot}$ of
iron group elements. It thus falls into the range of observational
expectations, albeit on the weaker side.\\

\begin{figure}[t!]
\centerline{
\includegraphics[width = 0.55 \linewidth]{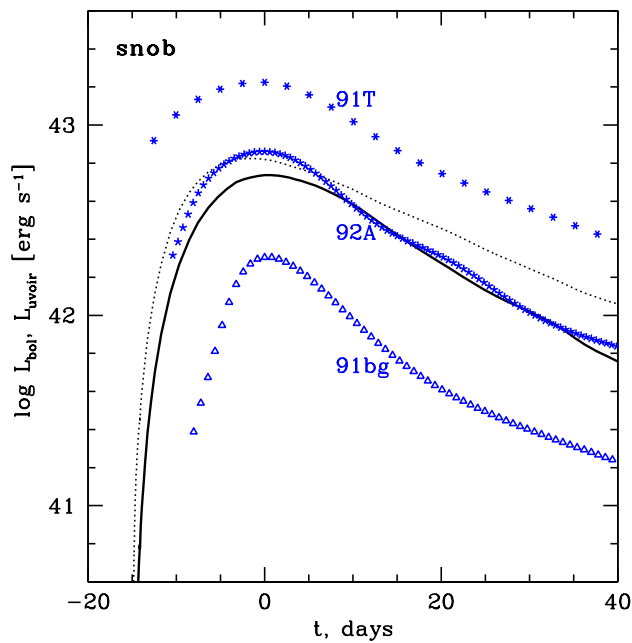}
}
\caption{Bolometric lightcurve derived from the highly resolved simulation (black curves;
  solid is the ``UVOIR-bolometric'' lightcurve and the complete
  bolometric lightcurve is dotted). The blue dotted curves correspond
  to observed bolometric lightcurves \citep{stritzinger2006a} (from \citep{roepke2007c}). \label{fig:lc}}
\end{figure}

\noindent \textbf{Lightcurves} ---
Lightcurves of SNe~Ia are sensitive to the energy release, the
$^{56}$Ni production, as well as to the distribution of elements in
the ejecta. In Figure~\ref{fig:lc} a synthetic bolometric lightcurve
derived from the above described simulation is compared with observed
lightcurves \citep{roepke2007b}. It was  
calculated using the \textsc{Stella} code
\citep{blinnikov1998a,blinnikov2000a,blinnikov2006a}. 
The synthetic bolometric lightcurve falls into the range of
observed lightcurves of SNe~Ia (indicated in Figure~\ref{fig:lc} by
the extreme examples of the super-luminous SN~1991T and the
sub-luminous SN~1991bg). With respect to brightness and shape it
compares reasonably well with SN~1994D, a standard normal SN~Ia.\\

\begin{figure}[t!]
\centerline{
\includegraphics[width = 0.63 \linewidth]{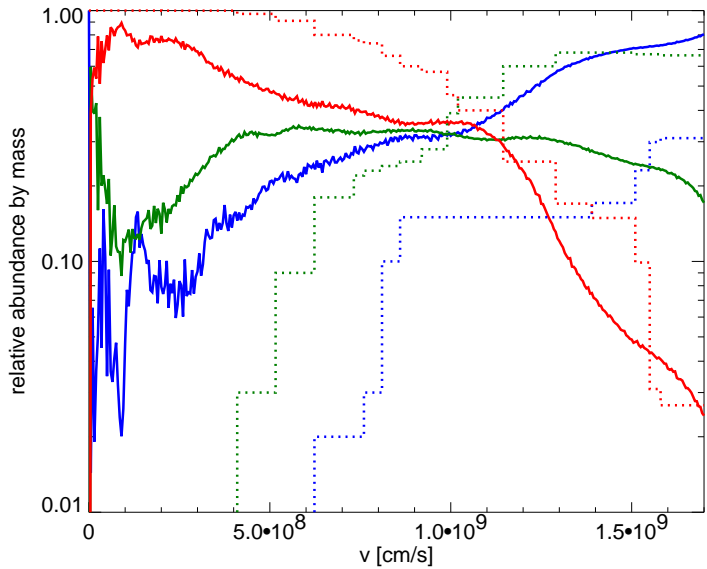}
}
\caption{Spherically averaged composition resulting from the
  hydrodynamical explosion simulation (solid lines) compared to the
  findings of the abundance tomography of SN~2002bo (dotted lines). Iron group
  element abundances are shown in red, intermediate-mass elements in
  green, and unburned material in blue (from \citep{roepke2007c}). \label{fig:at}}
\end{figure}

\noindent \textbf{Spectra} ---
A much harder test for the models is posed by the comparison of
synthetic and observed spectra since these depend on details in the
composition of the ejected material.
A powerful diagnostic tool to compare SN~Ia models with spectral observations
is provided by the abundance tomography of SN~2002bo
\citep{stehle2005a}. It makes 
use of spectra taken from this supernova with an extraordinary good time
coverage. Fitting this sequence of data with synthetic spectra
unveils the composition of the ejecta in velocity space slice by
slice (see Figure~\ref{fig:at}), since the photosphere moves gradually inwards with the expansion
of the remnant. 

This abundance tomography of the ejecta can be
compared with results of 3D models, when averaged
over the angles (Figure~\ref{fig:at}). Qualitatively, the mixed composition of the ejecta
found here is reproduced by deflagration SN~Ia
models in a natural way since these predict a distribution of burnt
material within the rising bubbles. A problem was, however,
that older models predicted large unburnt mass fractions in the central
parts of the ejecta in disagreement with the observational results
\citep{kozma2005a,stehle2005a}. The high-resolved simulation cures
this problem by clearly reproducing the iron-group dominance in the
low-velocity ejecta \citep{roepke2007c}. A good agreement between the
simulation and the abundance tomography of SN~2002bo is therefore
recovered in the inner parts of the ejecta. Above radii of
$\sim$$10,000\,\mathrm{km}\,\mathrm{s}^{-1}$, however, the simulation
predicts the chemical composition to be dominated by unburnt
material. This is in contradiction to observations which find the
material still to be dominated by nuclear ashes, mainly
intermediate-mass elements. It should however be noted that these
outer regions of the ejecta contain only little mass
($\sim$$0.25\,\mathrm{M}_{\odot}$) due to the low densities encountered here. Thus,
the observed chemical composition is reproduced by the model for the larger part of the
ejected material.

\subsection{Summary of the deflagration model}

The pure deflagration model has proven very successful in many
respects. It provides the up to now only fully self-consistent
thermonuclear supernova model. Apart from the initial conditions (flame
ignition, WD structure and composition), which are expected to vary in
nature, it describes the actual explosion process without tunable
parameters. Perhaps the most striking success is that it indeed leads
to an explosion of the WD releasing energies that fall into the range
of observational expectations. The structure of the ejecta seems to be
partially consistent with compositions of SN~Ia ejecta derived from
spectral observations. At least the inner part of the ejecta looks
similar to what is expected for the dimmer examples of normal SNe~Ia.

Despite these successes, there are shortcomings of the pure
deflagration model. Varying the initial conditions in a range that
seems physically plausible, it could not reproduce the full range of
observed SN~Ia diversity. In particular, the normal to bright examples
of the observational SN~Ia sample seem to be out of reach for the
deflagration model. For these to be reproduced, $^{56}$Ni masses of
the order of $0.7\, \mathrm{M}_{\odot}$ need to be synthesized in the explosion.
This could not be achieved thus far in the deflagration model. The
second concern is with regard to the composition of the ejecta. The
large-scale Rayleigh--Taylor instabilities lead to clumpy
inhomogeneities in the structure of the explosion ejecta and to a
strong mixing of different groups of reaction products in the angular
averaged profile. While the results may still be consistent with
observations in the central parts of the ejecta, the chemical
composition of the outer parts seems to be at odds with the
observation, where an intermediate-mass element dominance of the
ejecta is detected out to large radii. 
It seems that the burning provided by the deflagration flame is insufficient, in
particular in late phases of the explosion. Such an
incomplete burning would explain the shortcomings in reproducing the
outermost parts of the ejecta and the fact that the deflagration model
fails to explain the brighter SN~Ia events. Although the kinetic
energy of the explosion ejecta falls into the range of observational
expectations, it reaches only the lower end of what is expected for
normal SNe~Ia. 
Two interpretations are possible here:
\begin{enumerate}
\item The deflagration model accounts for a certain peculiar sub-class of
  SNe~Ia  \citep{phillips2007a}, namely rather dim events which show
  indications for unburnt material in their ejecta. 
\item The deflagration model is incomplete but provides an important
  building block for an extended explosion model. This would explain
  why it is rather successful in reproducing the dimmer SN~Ia
  events. For these, the deflagration phase may dominate the explosion
  process. 
\end{enumerate} 
In the following section, we elaborate on the second of these
possibilities. 

\section{The delayed detonation model}
\label{sect:dd}

\begin{figure}[t!]
\includegraphics[width=\textwidth]{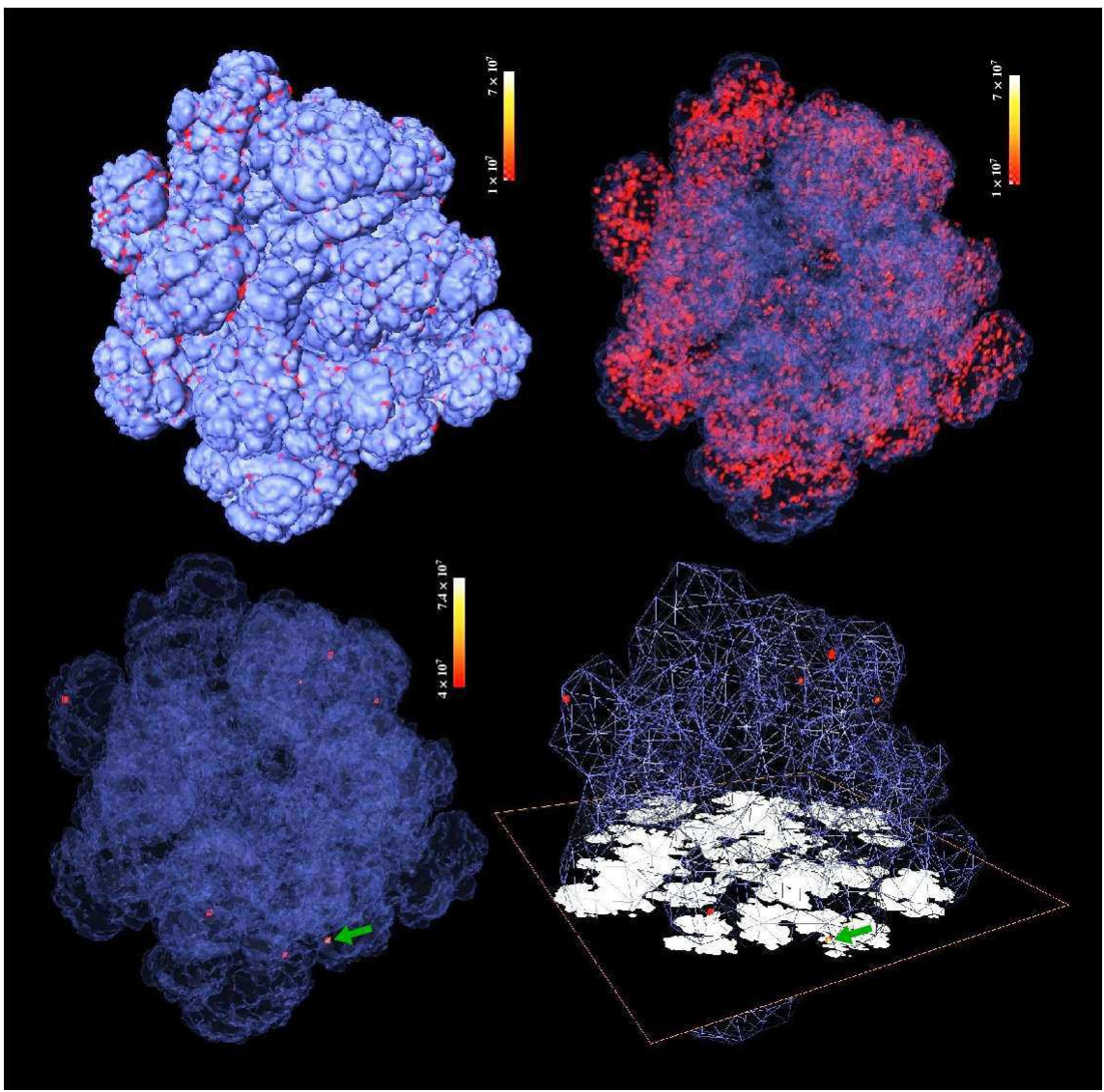}
\caption{Turbulent velocity fluctuations $v'$ at patches of the flame which
  have entered the distributed burning regime. The
  blue opaque, transparent, or wire-mesh surfaces correspond to the
  flame front. Volumes of high
  turbulent velocity fluctuations are rendered in red/orange. In the
  lower panels, the green arrow indicates the location of the maximum
  value of $v'$ found in the simulation. For better visibility, white
  areas correspond to ash regions and fuel regions are shown in black
  in a plane intersecting with the maximum $v'$-value in the lower
  right panel (from \citep{roepke2007d}). \label{fig:loc}}
\end{figure}

\begin{figure}[t!]
\centerline{\includegraphics[width=0.75\textwidth]{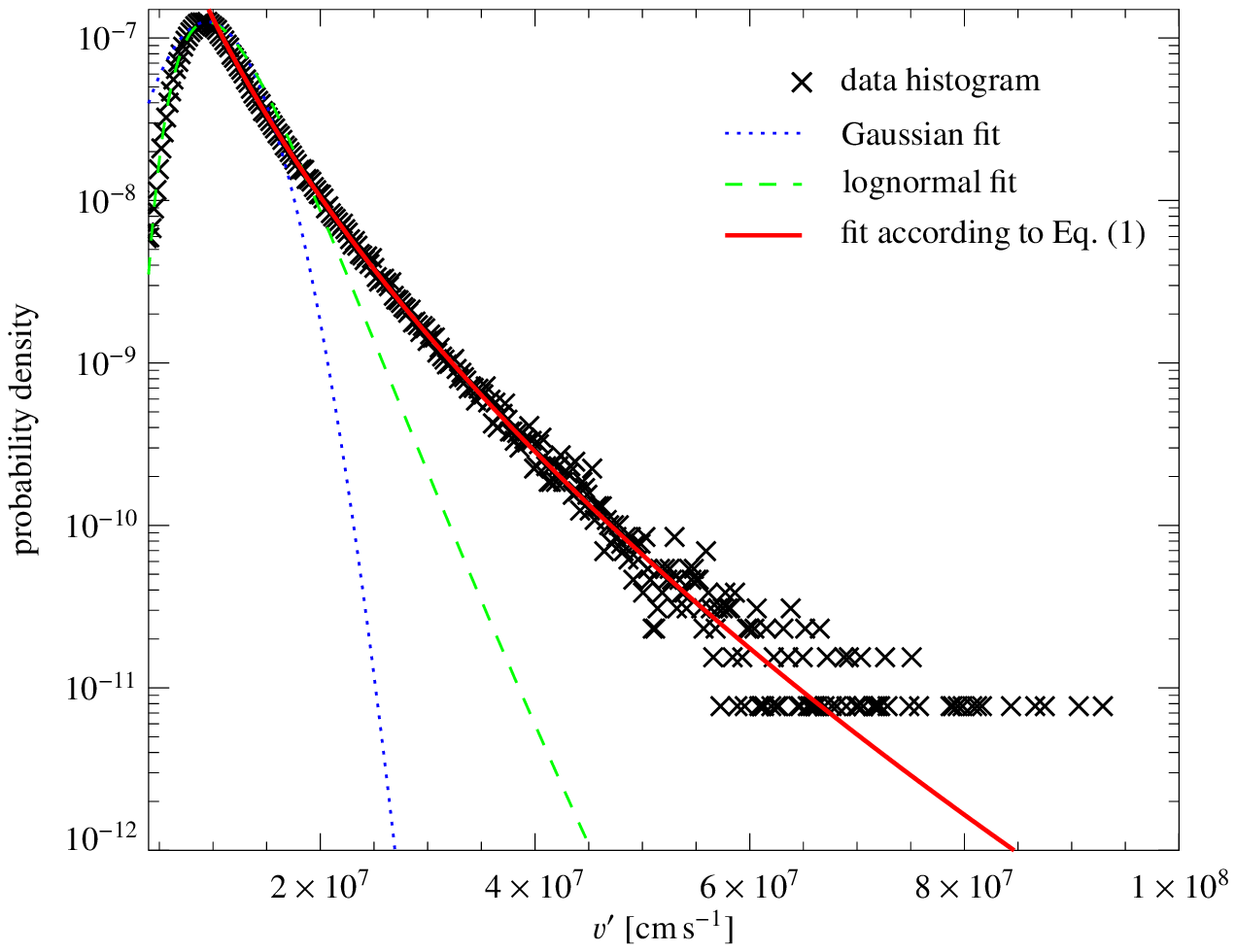}}
\caption{Fits to the histogram of the turbulent velocity fluctuations
  $v'$ (from \citep{roepke2007d}).\label{fig:fit}}
\end{figure}

A promising extension of the pure deflagration model is
the delayed detonation model \citep{khokhlov1991a}. By assuming a
transition of the flame propagation mode from subsonic deflagration to
supersonic detonation in an advanced stage of the explosion process,
it provides a way to accelerate and enhance burning in the late
phases. According to the discussion above, this may be a way to
achieve a better agreement with SN~Ia observations. Not only would it
lead to a stronger explosion and to an enrichment of the outer layers
of the ejecta with nuclear ashes, but it would also weaken the clumpy
inhomogeneities in the ejecta by supersonically burning down the
funnels in between the uprising Rayleigh--Taylor plumes.

\subsection{Deflagration-to-detonation transitions in SNe~Ia}

The delayed detonation model, however, is still afflicted with severe
uncertainties. Certainly the greatest hindrance for its success is
that the physical mechanism providing a deflagration-to-detonation
transition (DDT henceforth) in thermonuclear supernovae remains unclear
\citep{niemeyer1999a}. Such transitions are observed in terrestrial
combustion, but there they occur in the vicinity of obstacles or at
the walls
of the combustion vessel. Such boundaries do not exist in the
astrophysical situation. Therefore, it seems reasonable to assume that
a DDT occurs (if happening at all) in connection with the only
significant change in the structure of the deflagration flame that
occurs when it
enters the distributed burning regime
\citep{niemeyer1997b}. Remarkably, this takes place at fuel densities
of the order $10^7 \, \mathrm{g}\, \mathrm{cm}^{-3}$, coinciding with
the ``transition density'' that led to best results in parameterized
one-dimensional delayed detonation simulations.

DDTs occurring in
the distributed burning regime seem to be possible
\citep{lisewski2000b}, but only if a very high turbulence 
intensity is retained in these late explosion processes. Turbulent
velocity fluctuations of the order of $1000\, \mathrm{km} \,
\mathrm{s}^{-1}$ seem to be necessary to
trigger a DDT. While earlier
two-dimensional simulations of the deflagration phase in thermonuclear
supernovae failed to reach these values, a
recent analysis of three-dimensional simulations \citep{roepke2007d}
found that such high turbulent intensities are likely to be realized
in late burning stages. An example is given in Figure~\ref{fig:loc}
illustrating the situation in the highly resolved deflagration
simulation described in Sect.~\ref{sect:defl}.
Patches of the flame front are color-coded which have
entered the distributed burning regime and still feature high
turbulent velocities. A histogram of the velocity fluctuations
corresponding to the situation of Figure~\ref{fig:loc} is
shown in Figure~\ref{fig:fit}. It features an extended high-velocity
tail. An acceptable fit to this histogram is obtained with a
probability density function following an exponential of a geometric
Ansatz: 
\begin{equation}
\renewcommand{\theequation}{\arabic{equation}}
\label{eq:fit}
P(v') = \exp \left[ a_0 (v')^{a_1} +a_2  \right].
\end{equation}
For values of the fitting parameters $a_0$, $a_1$, and $a_2$ see \citep{roepke2007d}.
From such an analysis it is evident that the necessary high turbulent
velocity fluctuations are likely to be achieved at non-negligible
patches of the flame front and a flame-driven DDT thus cannot be
excluded. However, the microscopic mechanism of this transition
remains largely unclear. A high value of the mean turbulent
velocity is necessary, but certainly not sufficient
\citep{woosley2007a}. Further studies of the microscopic properties of
burning in the distributed regime are needed in order to settle the question
of a DDT in thermonuclear supernovae.

\subsection{Delayed detonations in full-star simulations}
\label{sect:dd_full_star}

Despite the uncertainties of DDTs, an
approach to address the question of delayed detonations in
thermonuclear supernovae is to \emph{assume} such a transition and to
artificially trigger the detonation wave in large simulations. The
question to be addressed by this approach is whether the outcome of
such models is consistent with observations. Prescribing the DDT to
occur at an arbitrary instant and a location near the center led to
promising results \citep{gamezo2005a}. However, such an approach
suffers from the arbitrariness of the assumed DDT and has little
predictive power. 

A way to assess the potential of the delayed detonation model is to fix
the DDT to some physically motivated guess
\citep{roepke2007b}. According to the discussion above,
one possibility is to assume 
the DDT to occur at the location where the flame first enters the
distributed burning regime \citep{golombek2005a,roepke2007b}. Although
this is a very simplifying assumption, it fixes the unknown DDT
parameter and allows to compare the results of the simulations with
observational expectations. This was done both in two-dimensional
\citep{golombek2005a} and three-dimensional \citep{roepke2007b}
simulations. Particular care was taken to prevent the detonation from
crossing ashes left behind by the previous deflagration stage. This
requirement follows from a recent analysis of the propagation of
detonations in WD matter \citep{maier2006a}. It can easily be
implemented by following the detonation wave with a level-set approach
\citep{golombek2005a, roepke2007a, fink2007a}.

\begin{figure}[t!]
\includegraphics[width=\textwidth]{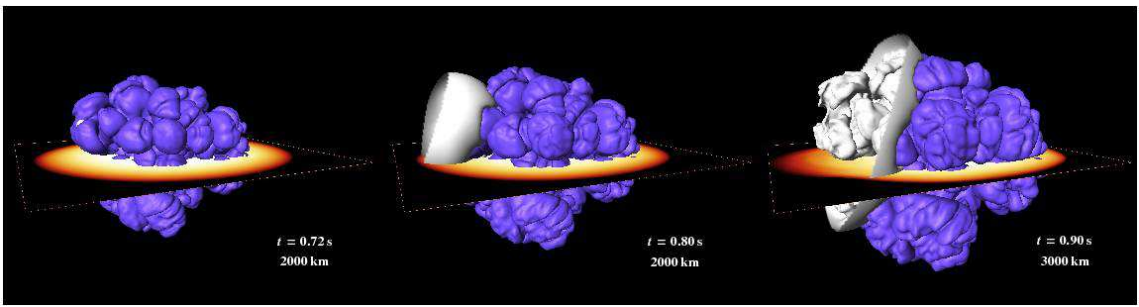}
\caption{Initiation and propagation of the detonation wave (white
  isosurface) in a delayed detonation model \citep{roepke2007b}. The deflagration flame is shown as
  blue isosurface and the extent of the star is indicated by the
  central plane mapping the logarithm of the density (from \citep{roepke2007b}).\label{fig:evo_dd}}
\end{figure}

A typical evolution of a three-dimensional full-star simulation is shown in
Figure~\ref{fig:evo_dd}. After ignition,
the deflagration phase proceeds in a way similar to the pure
deflagration model. The burning bubbles grow by flame propagation and
rise buoyantly towards the surface of the WD. Due to instabilities and
partial merger of the bubbles, a complex connected structure
develops. The parametrized DDT criterion is met first at $0.724 \,
\mathrm{s}$ after ignition at the outer edge of the flame front. Here,
the detonation is triggered by initiating the corresponding level set,
as shown in the left panel
of Fig.~\ref{fig:evo_dd}. The fact that the outer edge of the
deflagration flame is selected by the imposed DDT criterion is not
surprising. Turbulence is generated preferentially at large buoyant
bubbles and at the same time the density is lowest at the outermost
parts of the flame making the flame broadest here. This favors the
transition to the distributed burning regime and thus the parametrized
initiation of the detonation wave.

The center and right panels of Fig.~\ref{fig:evo_dd} show the
propagation of the detonation wave. Since it
cannot cross ash regions, it wraps around the corrugated
deflagration structure burning towards the star's center. In this way,
it takes about $0.2 \, 
\mathrm{s}$ before it arrives at the center of the WD. Meanwhile, the
star keeps expanding and the deflagration continues in regions not yet
reached by the detonation wave. Consequently, the density
of the fuel ahead of the detonation drops quickly after passing the WD's
center. This way, burning stalls shortly before the detonation
reaches the far side of the deflagration structure. 
Nonetheless, this still implies burning of most of the WD,
since these deflagration features have
already reached the low density edge of the star. 

Such simulations emphasize that the propagation of the detonation wave has
to compete with the expansion of the WD caused by the previous
deflagration phase. The outcome of this competition with respect to
the completeness of the burning, the $^{56}$Ni production (an thus the
brightness of the resulting event), as well as the energy release
depends strongly on the efficiency of burning in the deflagration
phase. This can be modified by varying the flame ignition geometry
\citep{roepke2007b} and leads to models that are capable of covering
the global characteristics of normal to bright observed SNe~Ia
\citep{roepke2007b}. Weak deflagrations lead to less expansion of the
WD before the detonation is triggered and the detonation thus
encounters a large amount of unburnt material at relatively high
densities. Therefore it significantly contributes to the overall
burning leading to very bright and energetic events dominated
by the detonation phase. In strong
deflagrations, in contrast, the high-density burning takes place
almost exclusively in the deflagration phase. The detonation thus
contributes little to the mass of radioactive nickel resulting in a
rather dim and moderately energetic explosion which bears the imprint
of the deflagration phase. Interestingly, however, the detonation at
lower densities leads to a layer of intermediate mass elements
surrounding the iron-group rich ejecta. Changing the flame ignition
(for instance 
by assuming a multi-spot ignition scenario \citep{roepke2006a,
  schmidt2006a}) it is possible to smoothly shift the characteristics
of the result from a deflagration-type explosion to a
detonation-dominated event.

\section{Conclusions}

\begin{figure}[t!]
\centerline{\includegraphics[width=0.66\textwidth]{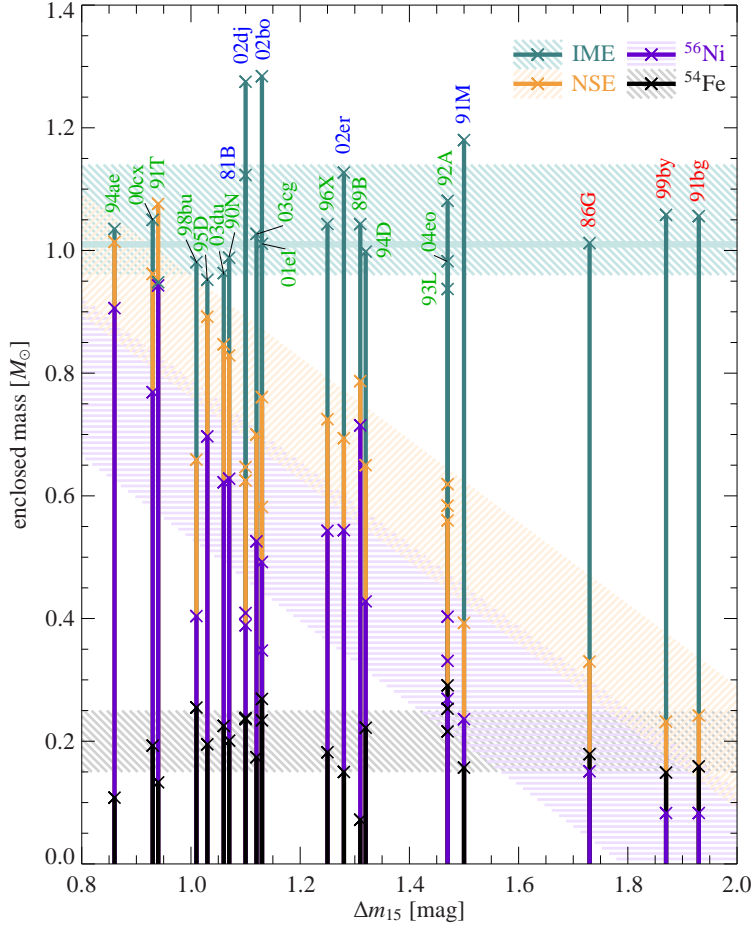}}
\caption{Distribution of the
  principal isotopic groups in SNe~Ia. The enclosed mass (linked to
  velocity via the W7 explosion model) of different burning products
  is shown versus decline-rate parameter $\Delta m_{15} (B)$ (a proxy
  for SN luminosity). 
  The mass of stable $^{54}$Fe+$^{58}$Ni for each SN is indicated in black; 
  that of $^{56}$Ni is shown in purple, and the sum of these (total NSE
  mass) is indicated in orange colors. Turquoise crosses show the sum
  of NSE and IME mass, indicating the 
  total mass burned. The IME mass is given in turquoise color (from \citep{mazzali2007a}).\label{fig:zorro}}
\end{figure}

As a conclusion, a speculation on the overall picture of thermonuclear
supernovae as a model of SNe~Ia can be provided based on \citep{mazzali2007a}. Spectra of a set of
well-observed SNe~Ia allowed to derive the composition of the ejecta
of the events. The result is shown in
Figure~\ref{fig:zorro} and indicates that the mass of stable iron group
elements is roughly constant for all events ($\sim$$0.2\,
\mathrm{M}_{\odot}$). The mass of radioactive $^{56}$Ni determines (as expected)
the brightness, and the total mass of iron group elements shows a
linear correlation with brighness. Interestingly, in all supernovae
roughly the same total mass of burnt material is found, i.e.\ the
dimmer events compensate less iron group elements in the ejecta
by a larger mass of intermediate-mass elements.

In the theoretical picture outlined above, a possible interpretation
is that the dimmer examples of the normal events (corresponding to the
central part of the diagram in Figure~\ref{fig:zorro} around $\Delta
m_{15} (B) \sim 1.5 \, \mathrm{mag}$) could be accounted for by pure
deflagrations. However, because of the difficulty to synthesize
large amounts of intermediate mass 
elements in the outer layers of the ejecta in
deflagration models, an alternative interpretation in the context
of delayed detonations seems more realistic. As discussed in
Sect.~\ref{sect:dd_full_star}, changing the flame ignition configuration
these models facilitates a smooth shift from the characteristics of
the model being dominated by the deflagration phase to a detonation-dominated
event. In this picture, the SNe~Ia in the central part of the diagram
in Figure~\ref{fig:zorro} could be explained by the weaker and
dimmer models with emphasis on the deflagration phase while the
detonation-dominated models seem promising for
reproducing the bright and energetic events (on the left-hand side of
the plot in Figure~\ref{fig:zorro}). In this hypothetic picture,
however, the peculiar sub-luminous SNe~Ia located to the right in the
plot remain unexplained. Further modeling efforts and a detailed
derivation of synthetic observables from the results of numerical
simulations will provide a way of testing this conjectured
scenario. Alternative thermonuclear supernova models, such as
gravitationally confined detonations and pulsational reverse
detonations, are examined on the basis of comparison with observations
as well. While the former scenario seems to generically lead to very
bright events only, a potential problem of the latter model is the
large amount of 
iron-group elements in the outer layer of the ejecta
\citep{baron2008a}. Clearly, further exploration seems necessary to settle
the question whether these models can reproduce normal SNe~Ia or lead
to peculiar events.

\begin{acknowledgments}
I would like to thank the organizers of the conference for the kind
invitation. I enjoyed the inspiring atmosphere at the conference.
\end{acknowledgments}


\begin{thebibliography}{10}
\providecommand{\url}[1]{\texttt{#1}}
\providecommand{\urlprefix}{URL }
\providecommand{\eprint}[2][]{\url{#2}}

\bibitem{francois2004a}
P.~{Fran{\c c}ois}, F.~{Matteucci}, R.~{Cayrel}, M.~{Spite}, F.~{Spite} et~al.
\newblock \textit{The evolution of the Milky Way from its earliest phases:
  {C}onstraints on stellar nucleosynthesis}.
\newblock \textit{\aap}, \textbf{421} (2004) 613.
\newblock \href{http://arXiv.org/abs/arXiv:astro-ph/0401499}{{[\tt
  arXiv:astro-ph/0401499]}}.

\bibitem{branch1992a}
D.~{Branch} and G.~A. {Tammann}.
\newblock \textit{Type {Ia} supernovae as standard candles}.
\newblock \textit{\araa}, \textbf{30} (1992) 359.

\bibitem{riess1998a}
A.~G. {Riess}, A.~V. {Filippenko}, P.~{Challis}, A.~{Clocchiatti}, A.~{Diercks}
  et~al.
\newblock \textit{Observational Evidence from Supernovae for an Accelerating
  Universe and a Cosmological Constant}.
\newblock \textit{\aj}, \textbf{116} (1998) 1009.
\newblock \href{http://arXiv.org/abs/arXiv:astro-ph/9805201}{{[\tt
  arXiv:astro-ph/9805201]}}.

\bibitem{perlmutter1999a}
S.~{Perlmutter}, G.~{Aldering}, G.~{Goldhaber}, R.~A. {Knop}, P.~{Nugent}
  et~al.
\newblock \textit{Measurements of {O}mega and {L}ambda from 42 High-Redshift
  Supernovae}.
\newblock \textit{\apj}, \textbf{517} (1999) 565.
\newblock \href{http://arXiv.org/abs/arXiv:astro-ph/9812133}{{[\tt
  arXiv:astro-ph/9812133]}}.

\bibitem{leibundgut2001a}
B.~{Leibundgut}.
\newblock \textit{Cosmological Implications from Observations of {T}ype {Ia}
  Supernovae}.
\newblock \textit{\araa}, \textbf{39} (2001) 67.

\bibitem{astier2005a}
P.~{Astier}, J.~{Guy}, N.~{Regnault}, R.~{Pain}, E.~{Aubourg} et~al.
\newblock \textit{The {S}upernova {L}egacy {S}urvey: measurement of
  {$\Omega_{\mathrm{M}}$}, {$\Omega_{\Lambda}$} and {$w$} from the first year
  data set}.
\newblock \textit{\aap}, \textbf{447} (2006) 31.
\newblock \href{http://arXiv.org/abs/arXiv:astro-ph/0510447}{{[\tt
  arXiv:astro-ph/0510447]}}.

\bibitem{wood-vasey2007a}
W.~M. {Wood-Vasey}, G.~{Miknaitis}, C.~W. {Stubbs}, S.~{Jha}, A.~G. {Riess}
  et~al.
\newblock \textit{Observational Constraints on the Nature of {D}ark {E}nergy:
  {F}irst Cosmological Results from the {ESSENCE} Supernova Survey}.
\newblock \textit{\apj}, \textbf{666} (2007) 694.
\newblock \href{http://arXiv.org/abs/arXiv:astro-ph/0701041}{{[\tt
  arXiv:astro-ph/0701041]}}.

\bibitem{wheeler1990a}
J.~C. {Wheeler} and R.~P. {Harkness}.
\newblock \textit{Type {I} supernovae}.
\newblock \textit{Reports of Progress in Physics}, \textbf{53} (1990) 1467.

\bibitem{filippenko1997a}
A.~V. {Filippenko}.
\newblock \textit{Optical Spectra of Supernovae}.
\newblock \textit{\araa}, \textbf{35} (1997) 309.

\bibitem{zwicky1938a}
F.~{Zwicky}.
\newblock \textit{On Collapsed Neutron Stars}.
\newblock \textit{\apj}, \textbf{88} (1938) 522.

\bibitem{hoyle1960a}
F.~{Hoyle} and W.~A. {Fowler}.
\newblock \textit{Nucleosynthesis in Supernovae}.
\newblock \textit{\apj}, \textbf{132} (1960) 565.

\bibitem{truran1967a}
J.~W. {Truran}, W.~D. {Arnett} and A.~G.~W. {Cameron}.
\newblock \textit{Nucleosynthesis in supernova shock waves}.
\newblock \textit{Canadian Journal of Physics}, \textbf{45} (1967) 2315.

\bibitem{colgate1969a}
S.~A. {Colgate} and C.~{McKee}.
\newblock \textit{Early Supernova Luminosity}.
\newblock \textit{\apj}, \textbf{157} (1969) 623.

\bibitem{hillebrandt2000a}
W.~{Hillebrandt} and J.~C. {Niemeyer}.
\newblock \textit{Type {Ia} Supernova Explosion Models}.
\newblock \textit{\araa}, \textbf{38} (2000) 191.
\newblock \href{http://arXiv.org/abs/arXiv:astro-ph/0006305}{{[\tt
  arXiv:astro-ph/0006305]}}.

\bibitem{whelan1973a}
J.~{Whelan} and I.~J. {Iben}.
\newblock \textit{Binaries and Supernovae of Type {I}}.
\newblock \textit{\apj}, \textbf{186} (1973) 1007.

\bibitem{nomoto1982a}
K.~{Nomoto}.
\newblock \textit{Accreting white dwarf models for type {I} supernovae. {I}.
  Presupernova evolution and triggering mechanisms}.
\newblock \textit{\apj}, \textbf{253} (1982) 798.

\bibitem{iben1984a}
I.~{Iben}, Jr. and A.~V. {Tutukov}.
\newblock \textit{Supernovae of type {I} as end products of the evolution of
  binaries with components of moderate initial mass ({M} not greater than about
  9 solar masses)}.
\newblock \textit{\apjs}, \textbf{54} (1984) 335.

\bibitem{ruiz-lapuente2004a}
P.~{Ruiz-Lapuente}, F.~{Comeron}, J.~{M{\'e}ndez}, R.~{Canal}, S.~J. {Smartt}
  et~al.
\newblock \textit{The binary progenitor of {T}ycho {B}rahe's 1572 supernova}.
\newblock \textit{\nat}, \textbf{431} (2004) 1069.
\newblock \href{http://arXiv.org/abs/arXiv:astro-ph/0410673}{{[\tt
  arXiv:astro-ph/0410673]}}.

\bibitem{nomoto1985a}
K.~{Nomoto} and I.~{Iben}, Jr.
\newblock \textit{Carbon ignition in a rapidly accreting degenerate dwarf---{A}
  clue to the nature of the merging process in close binaries}.
\newblock \textit{\apj}, \textbf{297} (1985) 531.

\bibitem{garcia1995a}
D.~{Garcia-Senz} and S.~E. {Woosley}.
\newblock \textit{Type {Ia} Supernovae: {T}he Flame Is Born}.
\newblock \textit{\apj}, \textbf{454} (1995) 895.

\bibitem{woosley2004a}
S.~E. {Woosley}, S.~{Wunsch} and M.~{Kuhlen}.
\newblock \textit{Carbon Ignition in {T}ype {Ia} Supernovae: {A}n Analytic
  Model}.
\newblock \textit{\apj}, \textbf{607} (2004) 921.
\newblock \href{http://arXiv.org/abs/arXiv:astro-ph/0307565}{{[\tt
  arXiv:astro-ph/0307565]}}.

\bibitem{iapichino2006a}
L.~{Iapichino}, M.~{Br{\"u}ggen}, W.~{Hillebrandt} and J.~C. {Niemeyer}.
\newblock \textit{The ignition of thermonuclear flames in type {Ia}
  supernovae}.
\newblock \textit{\aap}, \textbf{450} (2006) 655.
\newblock \href{http://arXiv.org/abs/arXiv:astro-ph/0512300}{{[\tt
  arXiv:astro-ph/0512300]}}.

\bibitem{kuhlen2006a}
M.~{Kuhlen}, S.~E. {Woosley} and G.~A. {Glatzmaier}.
\newblock \textit{Carbon Ignition in {T}ype {Ia} Supernovae. {II}. {A}
  Three-dimensional Numerical Model}.
\newblock \textit{\apj}, \textbf{640} (2006) 407.
\newblock \href{http://arXiv.org/abs/arXiv:astro-ph/0509367}{{[\tt
  arXiv:astro-ph/0509367]}}.

\bibitem{hoeflich2002a}
P.~{H{\"o}flich} and J.~{Stein}.
\newblock \textit{On the Thermonuclear Runaway in {T}ype {Ia} Supernovae: {H}ow
  to Run Away?}
\newblock \textit{\apj}, \textbf{568} (2002) 779.
\newblock \href{http://arXiv.org/abs/arXiv:astro-ph/0104226}{{[\tt
  arXiv:astro-ph/0104226]}}.

\bibitem{roepke2006a}
F.~K. {R{\"o}pke}, W.~{Hillebrandt}, J.~C. {Niemeyer} and S.~E. {Woosley}.
\newblock \textit{Multi-spot ignition in type {Ia} supernova models}.
\newblock \textit{\aap}, \textbf{448} (2006) 1.
\newblock \href{http://arXiv.org/abs/arXiv:astro-ph/0510474}{{[\tt
  arXiv:astro-ph/0510474]}}.

\bibitem{schmidt2006a}
W.~{Schmidt} and J.~C. {Niemeyer}.
\newblock \textit{Thermonuclear supernova simulations with stochastic
  ignition}.
\newblock \textit{\aap}, \textbf{446} (2006) 627.
\newblock \href{http://arXiv.org/abs/arXiv:astro-ph/0510427}{{[\tt
  arXiv:astro-ph/0510427]}}.

\bibitem{plewa2004a}
T.~{Plewa}, A.~C. {Calder} and D.~Q. {Lamb}.
\newblock \textit{Type {Ia} Supernova Explosion: {G}ravitationally Confined
  Detonation}.
\newblock \textit{\apjl}, \textbf{612} (2004) L37.
\newblock \href{http://arXiv.org/abs/arXiv:astro-ph/0405163}{{[\tt
  arXiv:astro-ph/0405163]}}.

\bibitem{roepke2007a}
F.~K. {R{\"o}pke}, S.~E. {Woosley} and W.~{Hillebrandt}.
\newblock \textit{Off-Center Ignition in {T}ype {Ia} Supernovae. {I}. Initial
  Evolution and Implications for Delayed Detonation}.
\newblock \textit{\apj}, \textbf{660} (2007) 1344.
\newblock \href{http://arXiv.org/abs/arXiv:astro-ph/0609088}{{[\tt
  arXiv:astro-ph/0609088]}}.

\bibitem{sim2007a}
S.~A. {Sim}, D.~N. {Sauer}, F.~K. {R{\"o}pke} and W.~{Hillebrandt}.
\newblock \textit{Light curves for off-centre ignition models of {T}ype {Ia}
  supernovae}.
\newblock \textit{\mnras}, \textbf{378} (2007) 2.
\newblock \href{http://arXiv.org/abs/arXiv:astro-ph/0703764}{{[\tt
  arXiv:astro-ph/0703764]}}.

\bibitem{hillebrandt2007a}
W.~{Hillebrandt}, S.~A. {Sim} and F.~K. {R{\"o}pke}.
\newblock \textit{Off-center explosions of {C}handrasekhar-mass white dwarfs:
  an explanation of super-bright type {Ia} supernovae?}
\newblock \textit{\aap}, \textbf{465} (2007) L17.
\newblock \href{http://arXiv.org/abs/arXiv:astro-ph/0702344}{{[\tt
  arXiv:astro-ph/0702344]}}.

\bibitem{arnett1969a}
W.~D. {Arnett}.
\newblock \textit{A Possible Model of Supernovae: Detonation of {$^{12}$C}}.
\newblock \textit{\apss}, \textbf{5} (1969) 180.

\bibitem{nomoto1976a}
K.~{Nomoto}, D.~{Sugimoto} and S.~{Neo}.
\newblock \textit{Carbon deflagration supernova, an alternative to carbon
  detonation}.
\newblock \textit{\apss}, \textbf{39} (1976) L37.

\bibitem{timmes1992a}
F.~X. {Timmes} and S.~E. {Woosley}.
\newblock \textit{The conductive propagation of nuclear flames. {I}. Degenerate
  {C+O} and {O+Ne+Mg} white dwarfs}.
\newblock \textit{\apj}, \textbf{396} (1992) 649.

\bibitem{ivanova1974a}
L.~N. {Ivanova}, V.~S. {Imshennik} and V.~M. {Chechetkin}.
\newblock \textit{Pulsation regime of the thermonuclear explosion of a star's
  dense carbon core}.
\newblock \textit{\apss}, \textbf{31} (1974) 497.

\bibitem{khokhlov1991a}
A.~M. {Khokhlov}.
\newblock \textit{Delayed detonation model for type {Ia} supernovae}.
\newblock \textit{\aap}, \textbf{245} (1991) 114.

\bibitem{woosley1994a}
S.~E. {Woosley} and T.~A. {Weaver}.
\newblock \textit{Massive stars, supernovae, and nucleosynthesis}.
\newblock In S.~A. {Bludman}, R.~{Mochkovitch} and J.~{Zinn-Justin}, eds.,
  \textit{Les Houches Session {LIV}: Supernovae}, pp. 63--154. North-Holland,
  Amsterdam, 1994.

\bibitem{arnett1994b}
D.~{Arnett} and E.~{Livne}.
\newblock \textit{The delayed-detonation model of {T}ype {Ia} supernovae. {II}.
  {T}he detonation phase}.
\newblock \textit{\apj}, \textbf{427} (1994) 330.

\bibitem{reinecke2002d}
M.~{Reinecke}, W.~{Hillebrandt} and J.~C. {Niemeyer}.
\newblock \textit{Three-dimensional simulations of type {Ia} supernovae}.
\newblock \textit{\aap}, \textbf{391} (2002) 1167.
\newblock \href{http://arXiv.org/abs/arXiv:astro-ph/0206459}{{[\tt
  arXiv:astro-ph/0206459]}}.

\bibitem{gamezo2003a}
V.~N. {Gamezo}, A.~M. {Khokhlov}, E.~S. {Oran}, A.~Y. {Chtchelkanova} and R.~O.
  {Rosenberg}.
\newblock \textit{Thermonuclear Supernovae: Simulations of the Deflagration
  Stage and Their Implications}.
\newblock \textit{Science}, \textbf{299} (2003) 77.
\newblock \href{http://arXiv.org/abs/arXiv:astro-ph/0212054}{{[\tt
  arXiv:astro-ph/0212054]}}.

\bibitem{roepke2006c}
F.~K. {R{\"o}pke}.
\newblock \textit{Multi-dimensional numerical simulations of type {Ia}
  supernova explosions}.
\newblock In S.~{R{\"o}ser}, ed., \textit{Reviews in Modern Astronomy},
  \textit{Reviews in Modern Astronomy}, vol.~19, pp. 127--156. 2006.

\bibitem{roepke2006d}
F.~K. {R{\"o}pke} and W.~{Hillebrandt}.
\newblock \textit{Three-dimensional Modeling of {T}ype {Ia} Supernova
  Explosions}.
\newblock In S.~{Kubono}, W.~{Aoki}, T.~{Kajino}, T.~{Motobayashi} and
  K.~{Nomoto}, eds., \textit{Origin of Matter and Evolution of Galaxies},
  \textit{American Institute of Physics Conference Series}, vol. 847, pp.
  190--195. 2006.

\bibitem{travaglio2004a}
C.~{Travaglio}, W.~{Hillebrandt}, M.~{Reinecke} and F.-K. {Thielemann}.
\newblock \textit{Nucleosynthesis in multi-dimensional {SN Ia} explosions}.
\newblock \textit{\aap}, \textbf{425} (2004) 1029.
\newblock \href{http://arXiv.org/abs/arXiv:astro-ph/0406281}{{[\tt
  arXiv:astro-ph/0406281]}}.

\bibitem{roepke2006b}
F.~K. {R{\"o}pke}, M.~{Gieseler}, M.~{Reinecke}, C.~{Travaglio} and
  W.~{Hillebrandt}.
\newblock \textit{Type {Ia} supernova diversity in three-dimensional models}.
\newblock \textit{\aap}, \textbf{453} (2006) 203.
\newblock \href{http://arXiv.org/abs/arXiv:astro-ph/0506107}{{[\tt
  arXiv:astro-ph/0506107]}}.

\bibitem{nomoto1984a}
K.~{Nomoto}, F.-K. {Thielemann} and K.~{Yokoi}.
\newblock \textit{Accreting white dwarf models of {T}ype {I} supernovae. {III}.
  Carbon deflagration supernovae}.
\newblock \textit{\apj}, \textbf{286} (1984) 644.

\bibitem{reinecke2002b}
M.~{Reinecke}, W.~{Hillebrandt} and J.~C. {Niemeyer}.
\newblock \textit{Refined numerical models for multidimensional type {Ia}
  supernova simulations}.
\newblock \textit{\aap}, \textbf{386} (2002) 936.
\newblock \href{http://arXiv.org/abs/arXiv:astro-ph/0111475}{{[\tt
  arXiv:astro-ph/0111475]}}.

\bibitem{roepke2005a}
F.~K. {R{\"o}pke} and W.~{Hillebrandt}.
\newblock \textit{The distributed burning regime in type {Ia} supernova
  models}.
\newblock \textit{\aap}, \textbf{429} (2005) L29.
\newblock \href{http://arXiv.org/abs/arXiv:astro-ph/0411667}{{[\tt
  arXiv:astro-ph/0411667]}}.

\bibitem{roepke2005b}
F.~K. {R{\"o}pke} and W.~{Hillebrandt}.
\newblock \textit{Full-star type {Ia} supernova explosion models}.
\newblock \textit{\aap}, \textbf{431} (2005) 635.
\newblock \href{http://arXiv.org/abs/arXiv:astro-ph/0409286}{{[\tt
  arXiv:astro-ph/0409286]}}.

\bibitem{roepke2007c}
F.~K. {R{\"o}pke}, W.~{Hillebrandt}, W.~{Schmidt}, J.~C. {Niemeyer}, S.~I.
  {Blinnikov} et~al.
\newblock \textit{A Three-Dimensional Deflagration Model for {T}ype {Ia}
  Supernovae Compared with Observations}.
\newblock \textit{\apj}, \textbf{668} (2007) 1132.
\newblock \href{http://arXiv.org/abs/arXiv:0707.1024}{{[\tt arXiv:0707.1024]}}.

\bibitem{roepke2004a}
F.~K. {R{\"o}pke}, W.~{Hillebrandt} and J.~C. {Niemeyer}.
\newblock \textit{The cellular burning regime in type {Ia} supernova
  explosions. {I}. {F}lame propagation into quiescent fuel}.
\newblock \textit{\aap}, \textbf{420} (2004) 411.
\newblock \href{http://arXiv.org/abs/arXiv:astro-ph/0312092}{{[\tt
  arXiv:astro-ph/0312092]}}.

\bibitem{roepke2006e}
F.~K. {R{\"o}pke}, W.~{Hillebrandt} and S.~I. {Blinnikov}.
\newblock \textit{On the Mechanism of {T}ype {Ia} Supernovae}.
\newblock In \textit{ESA Special Publication}, \textit{ESA Special
  Publication}, vol. 637. 2006.

\bibitem{gamezo2005a}
V.~N. {Gamezo}, A.~M. {Khokhlov} and E.~S. {Oran}.
\newblock \textit{Three-dimensional Delayed-Detonation Model of {T}ype {Ia}
  Supernovae}.
\newblock \textit{\apj}, \textbf{623} (2005) 337.
\newblock \href{http://arXiv.org/abs/arXiv:astro-ph/0409598}{{[\tt
  arXiv:astro-ph/0409598]}}.

\bibitem{golombek2005a}
I.~{Golombek} and J.~C. {Niemeyer}.
\newblock \textit{A model for multidimensional delayed detonations in {SN Ia}
  explosions}.
\newblock \textit{\aap}, \textbf{438} (2005) 611.
\newblock \href{http://arXiv.org/abs/arXiv:astro-ph/0503617}{{[\tt
  arXiv:astro-ph/0503617]}}.

\bibitem{roepke2007b}
F.~K. {R{\"o}pke} and J.~C. {Niemeyer}.
\newblock \textit{Delayed detonations in full-star models of type {Ia}
  supernova explosions}.
\newblock \textit{\aap}, \textbf{464} (2007) 683.

\bibitem{bravo2007a}
E.~{Bravo} and D.~{Garcia-Senz}.
\newblock \textit{A Three-Dimensional Picture of the Delayed-Detonation Model
  of {T}ype {Ia} Supernovae}.
\newblock \textit{ArXiv e-prints}, \textbf{712} (2007).
\newblock \href{http://arXiv.org/abs/arXiv:0712.0510}{{[\tt arXiv:0712.0510]}}.

\bibitem{plewa2007a}
T.~{Plewa}.
\newblock \textit{Detonating Failed Deflagration Model of Thermonuclear
  Supernovae. {I.} {E}xplosion Dynamics}.
\newblock \textit{\apj}, \textbf{657} (2007) 942.
\newblock \href{http://arXiv.org/abs/arXiv:astro-ph/0611776}{{[\tt
  arXiv:astro-ph/0611776]}}.

\bibitem{townsley2007a}
D.~M. {Townsley}, A.~C. {Calder}, S.~M. {Asida}, I.~R. {Seitenzahl}, F.~{Peng}
  et~al.
\newblock \textit{Flame Evolution During {T}ype {Ia} Supernovae and the
  Deflagration Phase in the Gravitationally Confined Detonation Scenario}.
\newblock \textit{\apj}, \textbf{668} (2007) 1118.
\newblock \href{http://arXiv.org/abs/arXiv:0706.1094}{{[\tt arXiv:0706.1094]}}.

\bibitem{roepke2006f}
F.~K. {R{\"o}pke} and S.~E. {Woosley}.
\newblock \textit{Surface detonation in type {Ia} supernova explosions?}
\newblock \textit{Journal of Physics Conference Series}, \textbf{46} (2006)
  413.
\newblock \href{http://arXiv.org/abs/arXiv:astro-ph/0609691}{{[\tt
  arXiv:astro-ph/0609691]}}.

\bibitem{bravo2006a}
E.~{Bravo} and D.~{Garc{\'{\i}}a-Senz}.
\newblock \textit{Beyond the Bubble Catastrophe of {T}ype {Ia} Supernovae:
  {P}ulsating Reverse Detonation Models}.
\newblock \textit{\apjl}, \textbf{642} (2006) L157.
\newblock \href{http://arXiv.org/abs/arXiv:astro-ph/0604025}{{[\tt
  arXiv:astro-ph/0604025]}}.

\bibitem{roepke2003a}
F.~K. {R{\"o}pke}, J.~C. {Niemeyer} and W.~{Hillebrandt}.
\newblock \textit{On the Small-Scale Stability of Thermonuclear Flames in
  {T}ype {Ia} Supernovae}.
\newblock \textit{\apj}, \textbf{588} (2003) 952.
\newblock \href{http://arXiv.org/abs/arXiv:astro-ph/0211202}{{[\tt
  arXiv:astro-ph/0211202]}}.

\bibitem{roepke2004b}
F.~K. {R{\"o}pke}, W.~{Hillebrandt} and J.~C. {Niemeyer}.
\newblock \textit{The cellular burning regime in type {Ia} supernova
  explosions. {II}. {F}lame propagation into vortical fuel}.
\newblock \textit{\aap}, \textbf{421} (2004) 783.
\newblock \href{http://arXiv.org/abs/arXiv:astro-ph/0312203}{{[\tt
  arXiv:astro-ph/0312203]}}.

\bibitem{zingale2005a}
M.~{Zingale}, S.~E. {Woosley}, C.~A. {Rendleman}, M.~S. {Day} and J.~B. {Bell}.
\newblock \textit{Three-dimensional Numerical Simulations of
  {R}ayleigh-{T}aylor Unstable Flames in {T}ype {Ia} Supernovae}.
\newblock \textit{\apj}, \textbf{632} (2005) 1021.
\newblock \href{http://arXiv.org/abs/arXiv:astro-ph/0501655}{{[\tt
  arXiv:astro-ph/0501655]}}.

\bibitem{schmidt2005e}
W.~{Schmidt}, W.~{Hillebrandt} and J.~C. {Niemeyer}.
\newblock \textit{Level set simulations of turbulent thermonuclear deflagration
  in degenerate carbon and oxygen}.
\newblock \textit{Combust.\ Theory Modelling}, \textbf{9} (2005) 693.
\newblock \href{http://arXiv.org/abs/arXiv:astro-ph/0508076}{{[\tt
  arXiv:astro-ph/0508076]}}.

\bibitem{peters2000a}
N.~{Peters}.
\newblock \textit{Turbulent Combustion}.
\newblock Cambridge University Press, Cambridge, 2000.

\bibitem{khokhlov1993a}
A.~{Khokhlov}.
\newblock \textit{Flame Modeling in Supernovae}.
\newblock \textit{\apjl}, \textbf{419} (1993) L77.

\bibitem{calder2004a}
A.~C. {Calder}, T.~{Plewa}, N.~{Vladimirova}, D.~Q. {Lamb} and J.~W. {Truran}.
\newblock \textit{Type {I}a Supernovae: An Asymmetric Deflagration Model}
  (2004).
\newblock ArXiv:astro-ph/0405126.

\bibitem{reinecke1999a}
M.~{Reinecke}, W.~{Hillebrandt}, J.~C. {Niemeyer}, R.~{Klein} and
  A.~{Gr{\"o}bl}.
\newblock \textit{A new model for deflagration fronts in reactive fluids}.
\newblock \textit{\aap}, \textbf{347} (1999) 724.
\newblock \href{http://arXiv.org/abs/arXiv:astro-ph/9812119}{{[\tt
  arXiv:astro-ph/9812119]}}.

\bibitem{osher1988a}
S.~{Osher} and J.~A. {Sethian}.
\newblock \textit{Fronts Propagating with Curvature-Dependent Speed:
  {A}lgorithms Based on {H}amilton--{J}acobi Formulations}.
\newblock \textit{Journal of Computational Physics}, \textbf{79} (1988) 12.

\bibitem{smiljanovski1997a}
V.~{Smiljanovski}, V.~{Moser} and R.~{Klein}.
\newblock \textit{A capturing--tracking hybrid scheme for deflagration
  discontinuities}.
\newblock \textit{Combustion Theory Modelling}, \textbf{1} (1997) 183.

\bibitem{hillebrandt2005a}
W.~{Hillebrandt}, M.~{Reinecke}, W.~{Schmidt}, F.~K. {R{\"o}pke},
  C.~{Travaglio} et~al.
\newblock \textit{Simulations of turbulent thermonuclear burning in {T}ype {Ia}
  supernovae}.
\newblock In G.~{Warnecke}, ed., \textit{Analysis and Numerics for Conservation
  Laws}, pp. 363--384. Springer, Berlin Heidelberg New York, 2005.
\newblock \href{http://arXiv.org/abs/arXiv:astro-ph/0405209}{{[\tt
  arXiv:astro-ph/0405209]}}.

\bibitem{roepke2004c}
F.~K. {R{\"o}pke} and W.~{Hillebrandt}.
\newblock \textit{The case against the progenitor's carbon-to-oxygen ratio as a
  source of peak luminosity variations in type {Ia} supernovae}.
\newblock \textit{\aap}, \textbf{420} (2004) L1.
\newblock \href{http://arXiv.org/abs/arXiv:astro-ph/0403509}{{[\tt
  arXiv:astro-ph/0403509]}}.

\bibitem{roepke2005c}
F.~K. {R{\"o}pke}.
\newblock \textit{Following multi-dimensional type {Ia} supernova explosion
  models to homologous expansion}.
\newblock \textit{\aap}, \textbf{432} (2005) 969.
\newblock \href{http://arXiv.org/abs/arXiv:astro-ph/0408296}{{[\tt
  arXiv:astro-ph/0408296]}}.

\bibitem{schmidt2006c}
W.~{Schmidt}, J.~C. {Niemeyer}, W.~{Hillebrandt} and F.~K. {R{\"o}pke}.
\newblock \textit{A localised subgrid scale model for fluid dynamical
  simulations in astrophysics. {II}. {A}pplication to type {Ia} supernovae}.
\newblock \textit{\aap}, \textbf{450} (2006) 283.
\newblock \href{http://arXiv.org/abs/arXiv:astro-ph/0601500}{{[\tt
  arXiv:astro-ph/0601500]}}.

\bibitem{niemeyer1997d}
J.~C. {Niemeyer} and A.~R. {Kerstein}.
\newblock \textit{Burning regimes of nuclear flames in {SN Ia} explosions}.
\newblock \textit{New Astronomy}, \textbf{2} (1997) 239.

\bibitem{damkoehler1940a}
G.~{Damk{\"o}hler}.
\newblock \textit{{Der Einflu{\ss} der Turbulenz auf die Flammengeschwindigkeit
  in Gasgemischen}}.
\newblock \textit{Z. f. Elektroch.}, \textbf{46} (1940) 601.

\bibitem{niemeyer1995a}
J.~C. {Niemeyer} and W.~{Hillebrandt}.
\newblock \textit{Microscopic Instabilities of Nuclear Flames in Type {Ia}
  Supernovae}.
\newblock \textit{\apj}, \textbf{452} (1995) 779.

\bibitem{schmidt2007a}
W.~{Schmidt}.
\newblock \textit{On the applicability of the level set method beyond the
  flamelet regime in thermonuclear supernova simulations}.
\newblock \textit{\aap}, \textbf{465} (2007) 263.
\newblock \href{http://arXiv.org/abs/arXiv:astro-ph/0701416}{{[\tt
  arXiv:astro-ph/0701416]}}.

\bibitem{reinecke1999b}
M.~{Reinecke}, W.~{Hillebrandt} and J.~C. {Niemeyer}.
\newblock \textit{Thermonuclear explosions of {C}handrasekhar-mass {C+O} white
  dwarfs}.
\newblock \textit{\aap}, \textbf{347} (1999) 739.
\newblock \href{http://arXiv.org/abs/arXiv:astro-ph/9812120}{{[\tt
  arXiv:astro-ph/9812120]}}.

\bibitem{contardo2000a}
G.~{Contardo}, B.~{Leibundgut} and W.~D. {Vacca}.
\newblock \textit{Epochs of maximum light and bolometric light curves of type
  {Ia} supernovae}.
\newblock \textit{\aap}, \textbf{359} (2000) 876.
\newblock \href{http://arXiv.org/abs/arXiv:astro-ph/0005507}{{[\tt
  arXiv:astro-ph/0005507]}}.

\bibitem{stritzinger2006a}
M.~{Stritzinger}, B.~{Leibundgut}, S.~{Walch} and G.~{Contardo}.
\newblock \textit{Constraints on the progenitor systems of type {Ia}
  supernovae}.
\newblock \textit{\aap}, \textbf{450} (2006) 241.
\newblock \href{http://arXiv.org/abs/arXiv:astro-ph/0506415}{{[\tt
  arXiv:astro-ph/0506415]}}.

\bibitem{blinnikov1998a}
S.~I. {Blinnikov}, R.~{Eastman}, O.~S. {Bartunov}, V.~A. {Popolitov} and S.~E.
  {Woosley}.
\newblock \textit{A Comparative Modeling of Supernova {1993J}}.
\newblock \textit{\apj}, \textbf{496} (1998) 454.
\newblock \href{http://arXiv.org/abs/arXiv:astro-ph/9711055}{{[\tt
  arXiv:astro-ph/9711055]}}.

\bibitem{blinnikov2000a}
S.~I. {Blinnikov} and E.~I. {Sorokina}.
\newblock \textit{{UV} light curves of thermonuclear supernovae}.
\newblock \textit{\aap}, \textbf{356} (2000) L30.
\newblock \href{http://arXiv.org/abs/arXiv:astro-ph/0003247}{{[\tt
  arXiv:astro-ph/0003247]}}.

\bibitem{blinnikov2006a}
S.~I. {Blinnikov}, F.~K. {R{\"o}pke}, E.~I. {Sorokina}, M.~{Gieseler},
  M.~{Reinecke} et~al.
\newblock \textit{Theoretical light curves for deflagration models of type {Ia}
  supernova}.
\newblock \textit{\aap}, \textbf{453} (2006) 229.
\newblock \href{http://arXiv.org/abs/arXiv:astro-ph/0603036}{{[\tt
  arXiv:astro-ph/0603036]}}.

\bibitem{stehle2005a}
M.~{Stehle}, P.~A. {Mazzali}, S.~{Benetti} and W.~{Hillebrandt}.
\newblock \textit{Abundance stratification in {T}ype {Ia} supernovae -- I. The
  case of {SN 2002bo}}.
\newblock \textit{\mnras}, \textbf{360} (2005) 1231.
\newblock \href{http://arXiv.org/abs/arXiv:astro-ph/0409342}{{[\tt
  arXiv:astro-ph/0409342]}}.

\bibitem{kozma2005a}
C.~{Kozma}, C.~{Fransson}, W.~{Hillebrandt}, C.~{Travaglio}, J.~{Sollerman}
  et~al.
\newblock \textit{Three-dimensional modeling of type {Ia} supernovae -- {T}he
  power of late time spectra}.
\newblock \textit{\aap}, \textbf{437} (2005) 983.
\newblock \href{http://arXiv.org/abs/arXiv:astro-ph/0504317}{{[\tt
  arXiv:astro-ph/0504317]}}.

\bibitem{phillips2007a}
M.~M. {Phillips}, W.~{Li}, J.~A. {Frieman}, S.~I. {Blinnikov}, D.~{DePoy}
  et~al.
\newblock \textit{The Peculiar {SN 2005hk}: {D}o Some {T}ype {Ia} Supernovae
  Explode as Deflagrations?}
\newblock \textit{\pasp}, \textbf{119} (2007) 360.
\newblock \href{http://arXiv.org/abs/arXiv:astro-ph/0611295}{{[\tt
  arXiv:astro-ph/0611295]}}.

\bibitem{roepke2007d}
F.~K. {R{\"o}pke}.
\newblock \textit{Flame-driven Deflagration-to-Detonation Transitions in {T}ype
  {Ia} Supernovae?}
\newblock \textit{\apj}, \textbf{668} (2007) 1103.
\newblock \href{http://arXiv.org/abs/arXiv:0709.4095}{{[\tt arXiv:0709.4095]}}.

\bibitem{niemeyer1999a}
J.~C. {Niemeyer}.
\newblock \textit{Can Deflagration-Detonation Transitions Occur in Type {Ia}
  Supernovae?}
\newblock \textit{\apjl}, \textbf{523} (1999) L57.
\newblock \href{http://arXiv.org/abs/arXiv:astro-ph/9906142}{{[\tt
  arXiv:astro-ph/9906142]}}.

\bibitem{niemeyer1997b}
J.~C. {Niemeyer} and S.~E. {Woosley}.
\newblock \textit{The Thermonuclear Explosion of {C}handrasekhar Mass White
  Dwarfs}.
\newblock \textit{\apj}, \textbf{475} (1997) 740.
\newblock \href{http://arXiv.org/abs/arXiv:astro-ph/9607032}{{[\tt
  arXiv:astro-ph/9607032]}}.

\bibitem{lisewski2000b}
A.~M. {Lisewski}, W.~{Hillebrandt} and S.~E. {Woosley}.
\newblock \textit{Constraints on the Delayed Transition to Detonation in {T}ype
  {Ia} Supernovae}.
\newblock \textit{\apj}, \textbf{538} (2000) 831.
\newblock \href{http://arXiv.org/abs/arXiv:astro-ph/9910056}{{[\tt
  arXiv:astro-ph/9910056]}}.

\bibitem{woosley2007a}
S.~E. {Woosley}.
\newblock \textit{Type {Ia} Supernovae: {B}urning and Detonation in the
  Distributed Regime}.
\newblock \textit{\apj}, \textbf{668} (2007) 1109.
\newblock \href{http://arXiv.org/abs/arXiv:0709.4237}{{[\tt arXiv:0709.4237]}}.

\bibitem{maier2006a}
A.~{Maier} and J.~C. {Niemeyer}.
\newblock \textit{{C+O} detonations in thermonuclear supernovae: interaction
  with previously burned material}.
\newblock \textit{\aap}, \textbf{451} (2006) 207.
\newblock \href{http://arXiv.org/abs/arXiv:astro-ph/0605293}{{[\tt
  arXiv:astro-ph/0605293]}}.

\bibitem{fink2007a}
M.~{Fink}, W.~{Hillebrandt} and F.~K. {R{\"o}pke}.
\newblock \textit{Double-detonation supernovae of sub-{C}handrasekhar mass
  white dwarfs}.
\newblock \textit{\aap}, \textbf{476} (2007) 1133.
\newblock \href{http://arXiv.org/abs/arXiv:0710.5486}{{[\tt arXiv:0710.5486]}}.

\bibitem{mazzali2007a}
P.~A. {Mazzali}, F.~K. {R{\"o}pke}, S.~{Benetti} and W.~{Hillebrandt}.
\newblock \textit{A Common Explosion Mechanism for {T}ype {Ia} Supernovae}.
\newblock \textit{Science}, \textbf{315} (2007) 825.
\newblock \href{http://arXiv.org/abs/arXiv:astro-ph/0702351}{{[\tt
  arXiv:astro-ph/0702351]}}.

\bibitem{baron2008a}
E.~{Baron}, D.~J. {Jeffery}, D.~{Branch}, E.~{Bravo}, D.~{Garc{\'{\i}}a-Senz}
  et~al.
\newblock \textit{Detailed Spectral Modeling of a Three-dimensional {P}ulsating
  {R}everse {D}etonation Model: {T}oo Much Nickel}.
\newblock \textit{\apj}, \textbf{672} (2008) 1038.
\newblock \href{http://arXiv.org/abs/arXiv:0709.4177}{{[\tt arXiv:0709.4177]}}.

\end{thebibliography}

%

\end{document}